\documentclass[aps,10pt,twocolumn,pra,superscriptaddress]{revtex4-1}
\usepackage{amssymb}

\usepackage{graphicx,color}
\usepackage{epstopdf}
\usepackage{epsfig}
\usepackage{mathrsfs}
\usepackage[breaklinks=true]{hyperref}

\begin{document}

\title{Generating Nonclassical Quantum Input Field States with Modulating Filters}
\author{John E.~Gough} \email{jug@aber.ac.uk}
\affiliation{Institute for Mathematics and Physics, Aberystwyth University,
SY23 3BZ, Wales, United Kingdom}

\author{Guofeng Zhang} \email{guofeng.zhang@polyu.edu.hk}
\affiliation{Department of Applied Mathematics, The Hong Kong Polytechnic
University, Hong Kong, China}
\date{\today}

\begin{abstract}
We give explicit constructions of quantum dynamical filters which generate
nonclassical states (coherent states, cat states, shaped single and
multi-photon states) of quantum optical fields as inputs to general quantum
Markov systems. The filters will be quantum harmonic oscillators damped by vacuum 
the input fields, and we exploit the fact that the cascaded filter and
system will have a Lindbladian that is naturally Wick-ordered in the filter
modes. In particular the initialization of the modulating filter will
determine the signal state generated.
\end{abstract}

\maketitle

\section{Introduction}

There has been considerable progress in the generation of nonclassical
states for continuous variable quantum fields such as shaped single and multiple photons \cite{L01}-\cite{E11}, cat states (superpositions of
coherent states) \cite{NN06}-\cite{O07a},etc., and this has been proposed for several quantum
technologies \cite{H02} - \cite{C97}. The issue of interest from a physical point of view is the
response of quantum open systems when the quantum input processes that drive them are prepared in
any of these non-classical field states

\begin{figure}[htbp]
\centering
\includegraphics[width=0.40\textwidth]{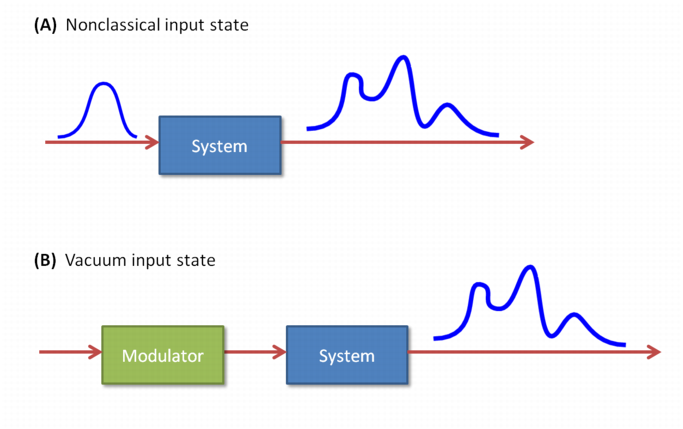}
\caption{(color online) System with nonclassical input; a coloring filter
(modulator) is used to convert vacuum noise input into a desired
non-classical input. Tracing out the modulating filter leads to the same effective
model.}
\label{fig:GP}
\end{figure}

Here we propose the use of quantum
mechanical modulating filters prepared in non-classical states which serve
to generate nonclassical quantum noise output from vacuum input, and that
may then be used to drive an open quantum system. This is an analogue of the 
techniques of coloring filters with have widespread applications in modeling classical 
engineering systems. The effect we seek is that, after tracing
out the modulator, the reduced quantum model will in effect be the one of the system driven
by an input field process in a nonclassical state, see Figure \ref{fig:GP}. The objective is to replace
the model of a system with non-classical noise with an effective model of 
a modulator driven by vacuum noise which is then cascaded with the system.
This opens up the possibility to access results from the established theory of
quantum input-outputs for initial vacuum inputs, and adapt them to non-classical
inputs.

\bigskip

\noindent \textbf{Notations:} Denote by $\mathscr{F}_{n}$ the span of all
symmetrized vectors of the form $f_{1}\hat{\otimes}\cdots \hat{\otimes}f_{n}=%
\frac{1}{n!}\sum_{\sigma }f_{\sigma (1)}\otimes \cdots \otimes f_{\sigma
(n)} $ where $f_{1},\cdots ,f_{n}$ lie in a one-particle Hilbert space $%
\mathscr{V}$, and the sum is over all permutations $\sigma $ of the $n$
indices. The Boson Fock space over $\mathscr{V}$ is then the direct sum $%
\mathscr{F}=\bigoplus_{n=0}^{\infty }\mathscr{F}_{n}$, with $\mathscr{F}_{0}$
spanned by the \textit{vacuum vector} $|$vac$\rangle $.

For $g \in \mathscr{V}$ and $T$ an operator on $\mathscr{V}$, the creation, annihilation and conservation operators are then given by ( $%
\widehat{f_{j}}$ indicating the omission of term $f_{j}$)
\begin{eqnarray*}
B\left( g\right) ^{\ast }f_{1}\hat{\otimes}\cdots \hat{\otimes}f_{n} &=&
\sqrt{n+1}g\hat{\otimes}f_{1}\hat{\otimes}\cdots \hat{\otimes}f_{n}, \\
B(g)f_{1}\hat{\otimes}\cdots \hat{\otimes}f_{n} &=&\frac{1}{\sqrt{n}}%
\sum_{j=1}^{n}\langle g|f_{j}\rangle \,f_{1}\hat{\otimes}\cdots \hat{\otimes}
\widehat{f_{j}}\hat{\otimes}\cdots \hat{\otimes}f_{n}, \\
\Lambda \left( T\right) f_{1}\hat{\otimes}\cdots \hat{\otimes}f_{n} &=&
\sum_{j=1}^{n}f_{1}\hat{\otimes}\cdots \hat{\otimes}(Tf_{j})\hat{\otimes}
\cdots \hat{\otimes}f_{n},
\end{eqnarray*}
and they map $\mathscr{F}_{n}$ to $\mathscr{F}_{n+1},\mathscr{F}_{n-1}$ and $%
\mathscr{F}_{n}$ respectively.

Given a complete orthonormal basis $\left\{ e_{1},e_{2},\cdots \right\} $
for $\mathscr{V}$, we obtain a complete orthonormal basis for $\mathscr{F}$
by setting
\begin{eqnarray*}
|\mathbf{n}\rangle =\sqrt{\frac{n!}{n_{1}!n_{2}!\cdots }}\widehat{\bigotimes
}_{k}e_{k}^{\otimes n_{k}} \equiv \prod_{k=1}^{\infty }\frac{1}{\sqrt{n_{k}!}%
}B\left( e_{k}\right) ^{\ast n_{k}}|\text{vac}\rangle
\end{eqnarray*}
where $\mathbf{n}=\left( n_{1},n_{2},\cdots \right) $ is a sequence of
occupation numbers and $n=\sum_{k}n_{k}$. The state $|\mathbf{n}\rangle $
corresponds to having $n_{k}$ photons in the state $e_{k}$ for each $k$.

For $\mathscr{V}=L^{2}[0,\infty )$, the space of square-integrable function $%
\xi (t)$ in $t \geq 0$, we denote the corresponding Fock space over $\mathscr{V}$ as $\mathfrak{F}$.
We then introduce the annihilation process $B_{t}=B(\chi
_{[ 0,t]})$ on $\mathfrak{F}$ where $\chi _{[ 0,t]}$ is the function equal to unity on the
interval 0 to $t$, and zero otherwise. The It\={o} differential $dB_{t}$ has
the action $dB_{t}|\mathbf{n}\rangle =\sum_{k=1}^{\infty }\sqrt{n_{k}}%
|n_{1},\cdots ,n_{k}-1,\cdots \rangle e_{k}(t)dt, $ for continuous test
functions $e_{k}$. For non-orthonormal states we have
\begin{eqnarray*}
dB_{t}\,f_{1}\hat{\otimes}\cdots \hat{\otimes}f_{n}=\frac{1}{\sqrt{n}}
\sum_{j}\,f_{1}\hat{\otimes}\cdots \hat{\otimes}\widehat{f_{j}}\hat{\otimes}
\cdots \hat{\otimes}f_{n}\,\,f_{j}(t)dt.
\end{eqnarray*}

For convenience we consider a single quantum input process.

Now fix a quantum mechanical system with Hilbert space $\mathfrak{h}_0$,
called the \textit{initial space}, then an open system is described by the
triple of operators $G\sim \left( S,L,H\right) $ on $\mathfrak{h}_0$ - with $%
S$ the unitary \textit{scattering matrix}, $L$ the \textit{collapse, or
coupling, operator} and $H$ the \textit{Hamiltonian} - which fixes the open
dynamical unitary evolution $U(t)$ on $\mathfrak{h}_0 \otimes \mathfrak{F}$
as the solution to the quantum stochastic differential equation \cite{HP}
\begin{eqnarray}
dU_{t}&=& \{ \left( S-I\right) \otimes d\Lambda _{t}+L\otimes dB_{t}^{\ast }
\nonumber \\
& & -L^{\ast }S\otimes dB_{t}- ( \frac{1}{2}L^{\ast }L+iH ) \otimes dt \} \,
U_{t}.  \label{eq:QSDE_U}
\end{eqnarray}
and in the Heisenberg picture we set $j_{t}(X)=U_{t}^{\ast }X \otimes I
U_{t} $ so that
\begin{eqnarray}
dj_{t}(X) &=& j_{t}(\mathcal{L}_{11 }X)\otimes d\Lambda _{t} +j_{t}(\mathcal{%
L}_{10 }X)\otimes dB^\ast_{t}  \nonumber \\
&&+j_{t}(\mathcal{L}_{01 }X)\otimes dB_{t} +j_{t}(\mathcal{L}_{00 }X)\otimes
d t
\end{eqnarray}
where we have the Evans-Hudson super-operators \cite{EH88}
\begin{eqnarray*}
\mathcal{L}_{00}X &=&\frac{1}{2}L^{\ast }[X,L]+\frac{1}{2}[L^{\ast },X%
]L-i[X,H], \\
\mathcal{L}_{10}X &=&S^{\ast }[X,L], \quad \mathcal{L}_{01}X = [L^{\ast
},X]S, \\
\mathcal{L}_{11}X &=&S^{\ast }XS-X.
\end{eqnarray*}

The \textit{output processes} are given by the formula
\begin{eqnarray}
B^{\mathrm{out}}_t = U^\ast(t) \, B(t) \, U(t),
\end{eqnarray}
and we have $dB^{\mathrm{out}}_t = j_t (S) dB_t + j_t (L) dt$.

Finally we recall that there is the natural factorization $\mathfrak{F} = %
\mathfrak{F}_t^- \otimes \mathfrak{F}^+_t$ of the Fock space into past and
future Fock spaces for each $t>0$ \cite{HP}.

\bigskip

\textbf{Definition 1 (from \cite{HP})} \textit{Let $\mathfrak{h}_1$ be a Hilbert space, and
 $X_{1}$ an operator on $\mathfrak{h}_{1}$, then given a second Hilbert space $\mathfrak{h}_2$
 we refer to the operator $X_{1}\otimes I_{2}$ as the ampliation of $X_1$ to the tensor
product space $\mathfrak{h}_{1}\otimes \mathfrak{h}_{2}$. A quantum stochastic process $(X(t))_{t\geq 0}$ is adapted if, for
each $t>0$, it is the ampliation of an operator on the past space $%
\mathfrak{h}_{0}\otimes \mathfrak{F}_{t}^{-}$ to the full space $\mathfrak{h}%
_{0}\otimes \mathfrak{F}$.}

\bigskip

The unitary evolution process $(U(t))_{t\geq 0}$ is adapted, as will be the
Heisenberg dynamical process $(j_t (X) )_{t \geq 0}$, for each initial
operator $X$. The following formula will be used extensively.

\bigskip

\textbf{Lemma} \textit{Let $(X(t))_{t\geq 0}$ be a quantum stochastic
integral process of the form}
\begin{eqnarray}
X(t) &=&X_{0}\otimes I_{\mathfrak{F}}+\int_{0}^{t}[x_{00}(s)ds  \nonumber \\
&&+x_{10}(s)dB_{s}^{\ast }+x_{01}(s)dB_{s}+x_{11}(s)d\Lambda _{s}],
\label{eq:QSIP}
\end{eqnarray}
\textit{where the $(x_{\alpha \beta }(t))_{t\geq 0}$ are adapted processes.
Then}
\begin{eqnarray}
&&d[U^{\ast }(t)X(t)U(t)]=  \nonumber \\
U^{\ast }(t)\bigg\{ &&[\mathcal{L}_{00}(X_{0})+x_{00}(t)+L^{\ast
}x_{10}(t)+x_{01}(t)L+x_{11}(t)L]dt  \nonumber \\
&&+[\mathcal{L}_{10}(X_{0})+S^{\ast }x_{10}(t)+S^{\ast }x_{11}(t)L]dB_{t}^{\ast
}  \nonumber \\
&&+[\mathcal{L}_{01}(X_{0})+x_{01}(t)S+L^{\ast }x_{11}(t)S]dB_{t}  \nonumber \\
&&+[\mathcal{L}_{11}(X_{0})+S^{\ast }x_{11}(t)S]d\Lambda _{t}\bigg\}U(t)
\label{eq:lemma}
\end{eqnarray}

The proof is a routine application of the quantum stochastic calculus \cite
{HP}. We note that if we set the $(x_{\alpha \beta } (t) )_{t \geq 0}$ equal
to zero, then we recover the Heisenberg equations of motion with initial
operators $X_0$. Conversely, setting $X_0=0$ and taking the $(x_{\alpha
\beta } (t) )_{t \geq 0}$ to be constants, the formula leads to the input-output
relation. Equation (\ref{eq:lemma}) therefore contains general information
about evolution of both system observables and field observables.

\section{Modulating Filter}

Our strategy is to employ a modulating filter $M$ to process vacuum input
and to feed this forward to the system. In principle, the modulator and
system are run in series as a single Markov component driven by vacuum
input, as in Figure \ref{fig:GP}. Tracing out the modulator degrees of
freedom leads to an effective model which leads to the same statistical
model as a non-vacuum input to the system. We shall show below how to
realize different non-classical driving fields in this way. In our proposal
we consider a linear passive system as modulator: physically corresponding
to modes in a cavity. The choice of (time-dependent) coupling operators
describing the modulator will be important in shaping the output, however,
in this set-up the crucial element determining non-vacuum statistics will be
the initial state $\phi _{0}\in \mathfrak{h}_{M}$ of the modulator.

We consider our system $G\sim \left( S,L,H\right) $ which is driven by the
output of a modulator $M\sim \left( I,L_{M},H_M\right) $ which itself is
driven by vacuum noise. The modulator and system in series is described by
the series product \cite{QFN1}, \cite{GJSeries} on the joint space $%
\mathfrak{h}_{M}\otimes \mathfrak{h}_{G}$ (here $\mathfrak{h}_G$ is the system Hilbert space
and $\mathfrak{h}_M$ is the modulator Hilbert space)
\begin{eqnarray}
\widetilde{G}&=& G\vartriangleleft M \sim ( I\otimes S,I\otimes L+L_{M}\otimes S,
\nonumber \\
&& I\otimes H+H_{M}\otimes I+\text{Im}\left\{ L_{M}\otimes L^{\ast
}S\right\} ) .
\end{eqnarray}

We note that the dynamical operators for the modulator are allowed to be time-varying, that is,
\[
L_M =L_M(t), \quad H_M=H_M(t),
\]
and indeed this will be a desirable feature for pulse shaping of the modulator output in what follows.
We suppress the $t$-dependence for notational convenience.

Let us denote by $\widetilde{U}_{t}$ the joint unitary generated by $\widetilde{G}$.
This is a unitary adapted process with initial space $\mathfrak{h}_0 = %
\mathfrak{h}_{M}\otimes \mathfrak{h}_{G}$.

\bigskip

\textbf{Definition 2} \textit{Let $G$ determine an open quantum system and
let $\Xi \in \mathfrak{F}$ be a state of the input field. A modulator $M$
with initial state $\phi _{0}$ and vacuum input is said to replicate the
open system if we have}
\begin{eqnarray}
&&\langle \phi _{0}\otimes \psi _{0}\otimes \mathrm{vac}|\widetilde{U}(t)^{\ast
}\,(I_{M}\otimes X(t))\,\widetilde{U}(t)|\phi _{0}\otimes \psi _{0}\otimes
\mathrm{vac}\rangle = \nonumber \\
&&\langle \psi _{0}\otimes \Xi |U(t)^{\ast }\,X(t)\,U(t)|\psi _{0}\otimes
\Xi \rangle
\label{eq:replicant}
\end{eqnarray}
\textit{for every adapted process $(X(t))_{t\geq 0}$ on $\mathfrak{h}%
_{G}\otimes \mathfrak{F}$ and all $\psi _{0}\in \mathfrak{h}_G$.}

\subsection{The Cascaded Lindbladian}

The total Lindbladian corresponding to $\widetilde{G}$ is
\begin{eqnarray}
\widetilde{\mathcal{L}}\left( A\otimes X\right) &=& \mathcal{L}_{M}\left(
A\right) \otimes X  \nonumber \\
&+& \sum_{\mu =0,1}\sum_{\nu =0,1}\left[ L_{M}^{\ast }\right] ^{\mu }A \left[
L_{M}\right] ^{\nu }\otimes \mathcal{L}_{\mu \nu }X ,
\label{eq:Lindblad_factor}
\end{eqnarray}
where $\mathcal{L}_{M}A=\frac{1}{2}\left[ L_{M}^{\ast },A\right] L_{M}+\frac{%
1}{2}L_{M}^{\ast }\left[ A,L_{M}\right] -i\left[ A,H_{M}\right] $ is the
modulator Lindbladian.

\subsection{Oscillator Mode Modulators}

Our interest will be in modulators that are linear passive systems. To this
end, we begin with the simplest model of a single Boson mode $a$ (say a
cavity mode) as modulator, and set
\begin{eqnarray}
H_{M}=\omega (t)\, a^{\ast }a,\quad L_{M}=\lambda (t) \, a.  \label{eq:osc}
\end{eqnarray}
For simplicity we shall take $\omega (t)\equiv 0$ and $\lambda $ to be a
complex-valued time-dependent damping parameter.

A key feature of equation (\ref{eq:Lindblad_factor}) when $L_{M}=\lambda
(t)a $ is that the $a$ and $a^{\ast }$ appear in \textit{Wick ordered form}
about $A$. We now exploit this property.

To compute expectations, we introduce the operator
\begin{eqnarray*}
\widetilde{a} _{t}\triangleq \widetilde{U}_{t}^{\ast }\left( a\otimes I\right)
\widetilde{U}_{t}
\end{eqnarray*}
and observe that $d\widetilde{a}_{t}=-z(t)\widetilde{a}_{t}\,dt-\lambda (t)^{\ast
}dB_{t}$ where we have the complex damping $z(t)=\frac{1}{2}\left| \lambda
(t)\right|^{2}+i\omega (t)$. The solution to this is the operator
\begin{eqnarray*}
\widetilde{a}_{t}=e^{-\zeta (t)}a-\int_{0}^{t}\lambda (s)^{\ast }e^{\zeta
(s)-\zeta (t)}dB_{s}
\end{eqnarray*}
with $\zeta (t)=\int_{0}^{t}z(s)ds$. Note that $\widetilde{a}_{t}$ consists of the ``deterministic term'' $e^{-\zeta (t)}a$ and a quantum stochastic integral term involving the field annihilator as integrator, so that we have
\[
\widetilde{a}_{t} | \phi_0 \otimes \psi_0 \otimes \mathrm{vac} \rangle = e^{-\zeta (t)}a| \phi_0 \otimes \psi_0 \otimes \mathrm{vac} \rangle .
\]
We get the following relation by virtue of the Wick ordering of the $a$ and $a^\ast$ terms, and the above identity.

\thinspace

\begin{widetext}
\begin{eqnarray}
&& \frac{d}{dt} \langle \phi_0 \otimes \psi_0 \otimes \mathrm{vac} |  \widetilde{U}(t)^\ast \,( I_M \otimes  X \otimes I_{\mathfrak{F}}  ) \, \widetilde{ U}(t) | \phi_0 \otimes \psi_0 \otimes \mathrm{vac} \rangle \nonumber \\
&=&
\sum_{\mu,\nu =0,1}
\langle \phi_0 \otimes \psi_0 \otimes \mathrm{vac} |
 \widetilde{U}(t)^\ast \,( \left[ \lambda \left( t\right) ^{\ast }a^{\ast }\right] ^{\mu } \left[
\lambda \left( t\right) a\right] ^{\nu }\otimes \mathcal{L}_{\mu \nu }X) \, \widetilde{ U}(t) | \phi_0 \otimes \psi_0 \otimes \mathrm{vac} \rangle \nonumber \\
&=&
\sum_{\mu,\nu =0,1}
\langle \phi_0 \otimes \psi_0 \otimes \mathrm{vac} |   \left[ \lambda \left( t\right) ^{\ast } \widetilde{a}_t^{\ast }\right] ^{\mu } \, \widetilde{U}(t)^\ast (I_M \otimes \mathcal{L}_{\mu \nu }X) \, \widetilde{ U}(t) \,  \left[
\lambda \left( t\right) \widetilde{a}_t\right] ^{\nu } | \phi_0 \otimes \psi_0 \otimes \mathrm{vac} \rangle \nonumber \\
&=& \sum_{\mu , \nu =0,1} \left[ \xi
(t)^{\ast }\right] ^{\mu }\left[ \xi \left( t\right) \right] ^{\nu }  \langle \left[ a\right] ^{\mu }\phi _{0}\otimes \psi _{0}\otimes
\text{vac}|  \widetilde{U}_{t}^{\ast }(I\otimes \mathcal{L}_{\mu \nu }X\otimes I)%
\widetilde{U}_{t}|\left[ a\right] ^{\nu }\phi _{0}\otimes \psi _{0}\otimes \text{%
vac}\rangle ,
\label{eq:main1}
\end{eqnarray}
\end{widetext}
where
\begin{equation}
\xi (t)=\lambda \left( t\right) e^{-\zeta (t)}.  \label{eq:xi}
\end{equation}

\subsection{Generating Shaped 1-Photon Fields}

The problem we ideally wish to solve is how to generate a desired pulse
shape $\xi $, and this means choosing the correct $\lambda $. It will be
required that $\xi $ be normalized, that is, $\int_{0}^{\infty }\left| \xi
(t)\right| ^{2}dt=1$. Now let us set $w(t)=\exp \left\{
-\int_{0}^{t}|\lambda (s)|^{2}ds\right\} $, then
\begin{eqnarray*}
\frac{d}{dt}w(t)=-|\lambda (t)|^{2}\exp \left\{ -\int_{0}^{t}|\lambda
(s)|^{2}ds\right\} \equiv -|\xi (t)|^{2}
\end{eqnarray*}
and, imposing the correct initial condition $w(0)=1$, we obtain
\begin{eqnarray}
w(t)\equiv \int_{t}^{\infty }\left| \xi (s)\right| ^{2}ds.  \label{ew:w}
\end{eqnarray}
Again taking $\omega (t)\equiv 0$ for simplicity, we find $z(t)\equiv \frac{1%
}{2}|\lambda (t)|^{2}$, real-valued. As we are given $\xi $ normalized, we
see that the appropriate choice for $\lambda $ is
\begin{eqnarray}
\lambda (t)=\frac{1}{\sqrt{w(t)}}\xi (t),  \label{eq:lambda}
\end{eqnarray}
with $w$ given by (\ref{ew:w}). An additional phase term will appear if we
have $\omega (t)$ non-zero.

\bigskip

\subsection{Replicating Non-vacuum Input}

Let $G\sim (S,L,H)$ and $M\sim \left( I_{M},\lambda a,\omega a^\ast a\right)
$ and let $\left( X\left( t\right) \right) _{t\geq 0}$ be a quantum
stochastic integral process on $\mathfrak{h}_{G}\otimes \mathfrak{F}$, as in
(\ref{eq:QSIP}) then we may generalize (\ref{eq:main1}) to get

$\quad$

\begin{widetext}
\begin{eqnarray}
&&\frac{d}{dt}\langle \phi _{0}\otimes \psi _{0}\otimes \mathrm{vac}|%
\widetilde{U}(t)^{\ast }\,(I_{M}\otimes X(t))\,\widetilde{U}(t)|\phi _{0}\otimes
\psi _{0}\otimes \mathrm{vac}\rangle  \nonumber \\
&=&\langle \phi _{0}\otimes \psi _{0}\otimes \mathrm{vac}|\widetilde{U}(t)^{\ast
}\,I_{M}\otimes \left( \mathcal{L}_{00}\left( X_{0}\right) +x_{00}+L^{\ast
}x_{10}+x_{01}L+L^{\ast }x_{11}L\right) \widetilde{U}(t)|\phi _{0}\otimes \psi
_{0}\otimes \mathrm{vac}\rangle  \nonumber \\
&&+\xi ^{\ast }\langle \phi _{0}\otimes \psi _{0}\otimes \mathrm{vac}%
|a^{\ast }\widetilde{U}(t)^{\ast }\,I_{M}\otimes \left( \mathcal{L}_{10}\left(
X_{0}\right) +S^{\ast }x_{10}+S^{\ast }x_{11}L\right) \widetilde{U}(t)|\phi
_{0}\otimes \psi _{0}\otimes \mathrm{vac}\rangle  \nonumber \\
&&+\xi\langle \phi _{0}\otimes \psi _{0}\otimes \mathrm{vac}|\widetilde{U}(t)^{\ast
}\,I_{M}\otimes \left( \mathcal{L}_{01}\left( X_{0}\right) +x_{01}S+L^{\ast
}x_{11}S\right) \widetilde{U}(t)a|\phi _{0}\otimes \psi _{0}\otimes \mathrm{vac}%
\rangle   \nonumber \\
&&+\xi ^{\ast }\xi \langle \phi _{0}\otimes \psi _{0}\otimes \mathrm{vac}%
|a^{\ast }\widetilde{U}(t)^{\ast }\,I_{M}\otimes \left( \mathcal{L}_{11}\left(
X_{0}\right) +S^{\ast }x_{11}S\right)
\widetilde{U}(t)a|\phi _{0}\otimes \psi _{0}\otimes \mathrm{vac}\rangle .
\label{eq:main_mod}
\end{eqnarray}
\end{widetext}

This follows from (\ref{eq:lemma}) where we replace the $S$ and $L$ with the
cascaded operators $I_{M}\otimes S$ and $I_{M}\otimes L+\lambda a\otimes S$.

For the modulator to replicate the dynamics with non-vacuum state $\Xi $,
the derivative
\[
\frac{d}{dt}\langle \psi _{0}\otimes \Xi |U\left( t\right) ^{\ast }X\left(
t\right) U\left( t\right) |\psi _{0}\otimes \Xi \rangle
\]
must equal the corresponding expression (\ref{eq:main_mod}) for any quantum stochastic integral process $X(t)$ on the Hilbert space $\mathfrak{h}_G \otimes \mathfrak{F}$. We may use (\ref{eq:lemma}) to show directly that this is computed from the following expectation

\thinspace

\begin{widetext}
\begin{eqnarray}
d\langle \psi _{0}\otimes \Xi |U\left( t\right) ^{\ast }X\left( t\right)
U\left( t\right) |\psi _{0}\otimes \Xi \rangle =\langle \psi _{0}\otimes \Xi
|U^{\ast }(t)\bigg\{ &&[\mathcal{L}_{00}(X_{0})+x_{00}(t)+L^{\ast
}x_{10}+x_{01}L+x_{11}L]dt  \nonumber \\
&&+[\mathcal{L}_{10}(X_{0})+S^{\ast }x_{10}(t)+S^{\ast }x_{11}L]dB_{t}^{\ast
}  \nonumber \\
&&+[\mathcal{L}_{01}(X_{0})+x_{01}(t)S+L^{\ast }x_{11}S]dB_{t}  \nonumber \\
&&+[\mathcal{L}_{11}(X_{0})+S^{\ast }x_{11}(t)S]d\Lambda _{t}\bigg\}%
U(t)|\psi _{0}\otimes \Xi \rangle   \label{eq:main_G}
\end{eqnarray}
\end{widetext}

The modulator therefore replicates the non-vacuum input model if (\ref{eq:main_G}) equals (\ref{eq:main_mod}).

\subsection{Replicating Coherent States}

As a simple illustration let us show how we may construct a modulator that
replicates a coherent state $|\beta \rangle $ for the input field, where $%
\beta \left( t\right) $ is a square integrable function of time $t\geq 0$.
Note that
\[
dB_{t}\,|\beta \rangle =\beta \left( t\right) \,|\beta \rangle \, dt .
\]
We see that the equations (\ref{eq:main_mod}) and (\ref{eq:main_G}) have
structural similarities, and a first guess for the initial state of the
modulator is another coherent state
\[
\phi _{0}=|\alpha \rangle
\]
where $\alpha \in \mathbb{C}$ is the intensity of the mode coherent state. In
this case $a\phi _{0}=\alpha \phi _{0}$ and (\ref{eq:main_mod}) and (\ref{eq:main_G}) coincide for the choice
\begin{eqnarray}
\xi \left( t\right) \alpha \equiv \beta \left( t\right) .
\label{eq:coherent_state_condition}
\end{eqnarray}
We therefore get the following result.

\bigskip

\textbf{Theorem 1:} \textit{The quantum open system }$G\sim \left(
S,L,H\right) $ \textit{driven by input in the continuous-variable coherent state }$\Xi
=\vert \beta \rangle $\textit{\ is replicated by the single mode modulator of the linear form}
$ M\sim \left( I_{M},\lambda (t) a,\omega (t) a^{\ast }a\right) $ \textit{ with the
initial state }$\phi _{0}=|\alpha \rangle $\textit{\ for the modulator and with }$%
\lambda \left( t\right) $\textit{\ and }$\omega \left( t\right) $\textit{\
chosen so that (\ref{eq:coherent_state_condition}) holds, for instance, if}
\begin{eqnarray}
\lambda (t) = \frac{ \beta (t)}{ \| \beta (t)  \sqrt{w(t)}}, \quad \omega (t) =0, \quad \alpha = \| \beta \| .
\end{eqnarray}

\section{Replicating Multi-photon Input}

\subsection{Fock State Input fields}

The state of a single mode quantum input field corresponding to $n$ quanta
with the same (normalized) one-particle test function $\xi \in
L^{2}[0,\infty )$ is
\[
\xi ^{\otimes n}=\frac{1}{\sqrt{n!}}B^{\ast }(\xi )^{n}|\text{vac}\rangle .
\]
We see that the annihilator acts on such states as
\[
dB_{t}\,\xi ^{\otimes n}=\sqrt{n}\xi (t)\xi ^{\otimes n-1}\,dt.
\]
One of the consequences of this comes when we try and compute expectations
of the form
\[
\langle \psi _{0}\otimes \xi ^{\otimes n}|x_{01}(t)dB_{t}|\psi _{0}\otimes
\xi ^{\otimes n}\rangle
\]
which becomes
\[
\sqrt{n}\xi (t)\,\langle \psi _{0}\otimes \xi ^{\otimes n}|x_{01}\left(
t\right) |\psi _{0}\otimes \xi ^{\otimes n-1}\rangle dt
\]
which is a matrix element between an $n$ photon state and an $n-1$ photon
state. This feature will be typical and so it is convenient to introduce
general matrix elements
\[
M_{t}^{n_{l},n_{r}}\left( X\right) =\langle \psi _{0}\otimes \xi ^{\otimes
n_{l}}|U(t)^{\ast }X\left( t\right) U\left( t\right) |\psi _{0}\otimes \xi
^{\otimes n_{r}}\rangle ,
\]
whenever $X\left( t\right) $ is a quantum stochastic integral and $%
n_{l},n_{r}\geq 0$. We set $M_{t}^{n_{l},n_{r}}\left( X\right) \equiv 0$ if
we ever have $n_{l}$ or $n_{r}$ negative.

Taking $X(t)$ to have the form (\ref{eq:QSIP}), we see that (\ref{eq:main_G})
leads to
\begin{eqnarray}
&&\frac{d}{dt}M_{t}^{n_{l},n_{r}}\left( X\right)   \nonumber \\
&=&M_{t}^{n_{l},n_{r}}\left( \mathcal{L}_{00}(X_{0})+x_{00}(t)+L^{\ast
}x_{10}+x_{01}L+x_{11}L\right)   \nonumber \\
&&+\sqrt{n}\xi ^{\ast }(t)\,M_{t}^{n_{l}-1,n_{r}}\left( \mathcal{L}%
_{10}(X_{0})+S^{\ast }x_{10}(t)+S^{\ast }x_{11}L\right)   \nonumber \\
&&+\sqrt{n}\xi (t)\,M_{t}^{n_{l},n_{r}-1}\left( \mathcal{L}%
_{01}(X_{0})+x_{01}(t)S+L^{\ast }x_{11}S\right)   \nonumber \\
&&+n\left| \xi (t)\right| ^{2}\,M_{t}^{n_{l}-1,n_{r}-1}\left( \mathcal{L}%
_{11}(X_{0})+S^{\ast }x_{11}(t)S\right) .  \label{eq:Fock_G}
\end{eqnarray}
We note the hierarchial nature of these equations with the rate of change of
$M_{t}^{n_{l},n_{r}}\left( X\right) $ depending on lower order matrix
elements.

Now let us introduce the single mode modulator, and let $|n\rangle $ be the
number states for the oscillator mode $\left( n=0,1,2,\cdots \right) $. We
may similarly introduce the matrix elements
\begin{eqnarray}
&&\widetilde{M}_{t}^{n_{l},n_{r}}\left( X\right) \nonumber \\
&=&\langle n_{l}\otimes \psi
_{0}\otimes \text{vac}|\widetilde{U}(t)^{\ast }(I_{M}\otimes X\left( t\right)
)U\left( t\right) |n_{r}\otimes \psi _{0}\otimes \text{vac}\rangle , \nonumber
\end{eqnarray}
with $n_{l},n_{r}\geq 0$, and $\widetilde{M}_{t}^{n_{l},n_{r}}\left( X\right) =0$
if either index is negative

From (\ref{eq:main_mod}), we see that
\begin{eqnarray}
&&\frac{d}{dt}\widetilde{M}_{t}^{n_{l},n_{r}}\left( X\right)   \nonumber \\
&=&\widetilde{M}_{t}^{n_{l},n_{r}}\left( \mathcal{L}_{00}(X_{0})+x_{00}(t)+L^{%
\ast }x_{10}+x_{01}L+x_{11}L\right)   \nonumber \\
&&+\sqrt{n}\xi ^{\ast }(t)\,\widetilde{M}_{t}^{n_{l}-1,n_{r}}\left( \mathcal{L}%
_{10}(X_{0})+S^{\ast }x_{10}(t)+S^{\ast }x_{11}L\right)   \nonumber \\
&&+\sqrt{n}\xi (t)\,\widetilde{M}_{t}^{n_{l},n_{r}-1}\left( \mathcal{L}%
_{01}(X_{0})+x_{01}(t)S+L^{\ast }x_{11}S\right)   \nonumber \\
&&+n\left| \xi (t)\right| ^{2}\,\widetilde{M}_{t}^{n_{l}-1,n_{r}-1}\left(
\mathcal{L}_{11}(X_{0})+S^{\ast }x_{11}(t)S\right) .  \label{eq:Fock_mod}
\end{eqnarray}

It follows that the systems of equations (\ref{eq:Fock_G}) and (\ref
{eq:Fock_mod}) are identical, and so we identify
\[
M_{t}^{n_{l},n_{r}}\left( X\right) \equiv \widetilde{M}_{t}^{n_{l},n_{r}}\left(
X\right)
\]
for all quantum stochastic integral process $X\left( t\right) $ on the joint
system and field. We summarize the result as follows.

\bigskip

\textbf{Theorem 2:} \textit{The quantum open system }$G\sim \left(
S,L,H\right) $ \textit{driven by input in the non-classical state }$\Xi
=\xi ^{\otimes n}$\textit{\ is replicated by the single mode modulator of the linear form}
$ M\sim \left( I_{M},\lambda (t)  a,\omega (t) a^{\ast }a\right) $ \textit{ with the
initial state }$\phi _{0}=|n\rangle $\textit{\ for the modulator and with }$%
\lambda \left( t\right) $\textit{\ and }$\omega \left( t\right) $\textit{\
chosen so that (\ref{eq:xi}) holds.}

For instance, we may again realize this with the specific choice $ \lambda (t)=\frac{1}{\sqrt{w(t)}}\xi (t)$, as in (\ref{eq:lambda}), and $\omega (t)=0$).

\subsection{General Multi-Photon Input Fields}

To generate multi-photon input field state (assumed normalized)
\begin{eqnarray}
\Xi (\mathbf{n})= \widehat{\bigotimes }_{k=1}^{N}\xi _{k}^{\otimes
n_{k}}.  \label{eq:general_state}
\end{eqnarray}
where now the $\xi_k$ are distinct, we need a multimode cavity with several
independent photon modes $a_{1},\cdots ,a_{N}$. The coupling operator may
now be extended to
\begin{eqnarray*}
L_{M}=\sum_{k}\lambda _{k}(t)a_{k}.
\end{eqnarray*}

It is convenient to introduce the vectors
\begin{eqnarray*}
\xi (t) &=& [ \xi_1 (t) , \cdots , \xi_N (t) ] \\
\lambda (t) &=& [ \lambda_1 (t) , \cdots , \lambda_N (t) ] \\
a &=&
\left[
\begin{array}{c}
a_1\\
\vdots \\
a_N
\end{array}
\right]
\end{eqnarray*}
so that $L_M \equiv \lambda (t) a$. We consider the vector of time-evolved modes $\widetilde{a}_t = \tilde{U}_t^\ast a \widetilde{U}_t$
and from the It\={o} rules, we find
\[
d \tilde{a}_t = A(t) \, \tilde{a}_t \ dt -\lambda (t)^\dag \, dB_t ,
\]
where $A(t)$ in the time-dependent $N\times N$ matrix with entries $A_{jk} (t) = - \frac{1}{2} \lambda_j^\ast (t) \lambda_k (t) -i \omega_k (t) \delta_{jk}$. The solution is
\[
\tilde{a}_t = \Phi (t) \, a - \Phi (t) \int_0^t \Phi(s)^{-1} \lambda (s)^\dag \, dB_s ,
\]
which is given in terms of the transition matrix $\Phi (t)$ satisfying
\[
\frac{d}{dt} \Phi (t) = A(t) \, \Phi (t) , \quad \Phi (0) = I_N.
\]
We then obtain the following vectorial generalization of (\ref{eq:main_mod})

$\quad$

\begin{widetext}
\begin{eqnarray}
&&\frac{d}{dt}\langle \phi _{0}\otimes \psi _{0}\otimes \mathrm{vac}|%
\widetilde{U}(t)^{\ast }\,(I_{M}\otimes X(t))\,\widetilde{U}(t)|\phi _{0}\otimes
\psi _{0}\otimes \mathrm{vac}\rangle  \nonumber \\
&=&\langle \phi _{0}\otimes \psi _{0}\otimes \mathrm{vac}|\widetilde{U}(t)^{\ast
}\,I_{M}\otimes \left( \mathcal{L}_{00}\left( X_{0}\right) +x_{00}+L^{\ast
}x_{10}+x_{01}L+L^{\ast }x_{11}L\right) \widetilde{U}(t)|\phi _{0}\otimes \psi
_{0}\otimes \mathrm{vac}\rangle  \nonumber \\
&&+ \langle \phi _{0}\otimes \psi _{0}\otimes \mathrm{vac}%
|a^{\ast }\Phi(t)^{\dag } \lambda (t)^\ast \widetilde{U}(t)^{\ast }\,I_{M}\otimes \left( \mathcal{L}_{10}\left(
X_{0}\right) +S^{\ast }x_{10}+S^{\ast }x_{11}L\right) \widetilde{U}(t)|\phi
_{0}\otimes \psi _{0}\otimes \mathrm{vac}\rangle  \nonumber \\
&&+ \langle \phi _{0}\otimes \psi _{0}\otimes \mathrm{vac}|\widetilde{U}(t)^{\ast
}\,I_{M}\otimes \left( \mathcal{L}_{01}\left( X_{0}\right) +x_{01}S+L^{\ast
}x_{11}S\right) \widetilde{U}(t) \lambda (t) \Phi (t) a|\phi _{0}\otimes \psi _{0}\otimes \mathrm{vac}%
\rangle   \nonumber \\
&&+ \langle \phi _{0}\otimes \psi _{0}\otimes \mathrm{vac}%
|a^{\ast } \Phi(t)^{\dag } \lambda (t)^\ast \widetilde{U}(t)^{\ast }\,I_{M}\otimes \left( \mathcal{L}_{11}\left(
X_{0}\right) +S^{\ast }x_{11}S\right)
\widetilde{U}(t) \lambda (t) \Phi (t ) a|\phi _{0}\otimes \psi _{0}\otimes \mathrm{vac}\rangle .
\label{eq:main_mod_vec}
\end{eqnarray}
\end{widetext}

Evidently, to get a prescribed set of pulses $\xi (t)$, we need to choose $\lambda (t)$ and the $\omega_k (t)$'s such that
\begin{eqnarray}
\xi (t) = \lambda (t) \, \Phi (t) .
\label{eq:Fock_multi}
\end{eqnarray}
In general this is a difficult problem to solve, but for weak pulses the Magnus expansion may offer a way to construct approximations.

We now prepare the modulator in the initial state
\begin{eqnarray}
\phi _{0}=|\mathbf{n}\rangle =|n_{1},\cdots ,n_{N}\rangle
\end{eqnarray}
where we have $n_{k}$ quanta in the $k$th cavity mode.

This time we consider the family of expectations
\begin{eqnarray*}
\widetilde{M}_{t}^{\mathbf{n}_{l},\mathbf{n}_{r}}( X)&=& \langle
\mathbf{n}_{l}\otimes \psi _{0}\otimes \text{vac}|  \nonumber \\
&& \widetilde{U}_{t}^{\ast }(I_M \otimes X(t) )\widetilde{U}_{t}|\mathbf{n}%
_{r}\otimes \psi _{0}\otimes \text{vac}\rangle ,
\end{eqnarray*}
for occupation sequences $\mathbf{n}_{l}=\left( n_{k,l}\right) $ and $%
\mathbf{n}_{r}=\left( n_{k,r}\right) $.

\begin{widetext}
\begin{eqnarray}
\frac{d}{dt}\widetilde{M}_{t}^{\mathbf{n}_{l},\mathbf{n}_{r}}  \left( X\right)
&=&\widetilde{M}_{t}^{\mathbf{n}_{l},\mathbf{n}_{r}}   \left( \mathcal{L}_{00}(X_{0})+x_{00}(t)+L^{%
\ast }x_{10}+x_{01}L+x_{11}L\right)   \nonumber \\
&&+ \sum_{k=1}^{N}\sqrt{n_{k,l}}\xi _{k}(t)^{\ast } \,\widetilde{M}_{t}^{\mathbf{n}_{l}-\mathbf{\delta}_k ,\mathbf{n}_{r}}  \left( \mathcal{L}%
_{10}(X_{0})+S^{\ast }x_{10}(t)+S^{\ast }x_{11}L\right)   \nonumber \\
&&+ \sum_{j=1}^{N}\sqrt{n_{j,r}}\xi _{j}(t) \,\widetilde{M}_{t}^{\mathbf{n}_{l},\mathbf{n}_{r}-\mathbf{\delta}_j} \left( \mathcal{L}%
_{01}(X_{0})+x_{01}(t)S+L^{\ast }x_{11}S\right)   \nonumber \\
&&+ \sum_{k=1}^{N}\sum_{j=1}^{N}\sqrt{n_{k,l}}\xi _{k}(t)^{\ast }\sqrt{n_{j,r}}\xi _{j}(t) \,\widetilde{M}_{t}^{\mathbf{n}_{l}-\mathbf{\delta}_k,\mathbf{n}_{r}-\mathbf{\delta}_j}\left(
\mathcal{L}_{11}(X_{0})+S^{\ast }x_{11}(t)S\right) ,
\label{eq:Fock_mod_vec}
\end{eqnarray}
\end{widetext}

where now $\mathbf{\delta }_{k}$ is the occupation sequence where $n_{k}=1$
and all other terms are zero. We add sequences together in the obvious way
so that $\mathbf{n}-\mu \mathbf{\delta }_{k}$ equals $\mathbf{n}$ if $\mu =0$%
, and $(n_{1},\cdots ,n_{k}-1,\cdots ,n_{N})$ if $\mu =1$.

By similar arguments as before, we see that system of expectations
\begin{eqnarray*}
M_{t}^{\mathbf{n}_{l},\mathbf{n}_{r}}( X)&=& \langle
 \psi _{0}\otimes \Xi (\mathbf{n}_l )|  U_{t}^{\ast }  X(t) U_{t}|  \psi _{0}\otimes \Xi (\mathbf{n}_r )\rangle 
\end{eqnarray*}
generate the same system of as the $\widetilde{M}_{t}^{\mathbf{n}_{l},\mathbf{n}_{r}}( X)$ and so may be equated.
We have therefore established that

\bigskip

\textbf{Theorem 3:} \textit{The quantum open system }$G\sim \left(
S,L,H\right) $ \textit{driven by input in the non-classical state }$\Xi (\mathbf{n})=\widehat{\bigotimes }_{k=1}^{N}\xi _{k}^{\otimes
n_{k}} $\textit{\ is replicated by the $N$ mode modulator of the linear form}
$ M\sim \left( I_{M},\sum_k \lambda_k (t)  a,\sum_k \omega_k (t)  a_k^{\ast }a_k\right) $ \textit{ with the
initial state }$\phi _{0}=|\mathbf{n} \rangle $\textit{\ for the modulator and with }$%
\lambda_k \left( t\right) $\textit{\ and }$\omega_k \left( t\right) $\textit{\
chosen so that (\ref{eq:Fock_multi}) holds.}

\section{Superposition Principles}

We now make a basic observation.

\bigskip

\textbf{Principle of Superimposed models:}
\textit{
For a fixed modulator $M$ - that is, a quantum open system with definite $(I_M,L_M,H_M)$ - suppose that initial states $
| \phi_0^A \rangle , | \phi_0^B \rangle , \cdots$ replicate $| \Xi^A \rangle , |\Xi^B \rangle , \cdots$ respectively and are compatible in so far as
\begin{eqnarray}
&&\langle \phi _{0}^A \otimes \psi _{0} \otimes \mathrm{vac}|\widetilde{U}(t)^{\ast
}\,(I_{M}\otimes X(t))\,\widetilde{U}(t)|\phi _{0}^B\otimes \psi _{0}\otimes
\mathrm{vac}\rangle = \nonumber \\
&&\langle \psi _{0}\otimes \Xi^A |U(t)^{\ast }\,X(t)\,U(t)|\psi _{0}\otimes
\Xi^B \rangle
\label{eq:compatible_replicant}
\end{eqnarray}
for each pair $A,B$.
Then if the modulator is prepared in a normalized superposition $|\phi_0 \rangle = c_A | \phi_0^A \rangle + c_B  | \phi_0^B \rangle + \cdots$
it will then replicate the nonclassical state $| \Xi \rangle = c_A | \Xi^A \rangle + c_B  | \Xi^B \rangle + \cdots$.}

\bigskip

This follows automatically from the bra-ket structure of the matrix elements.

\subsection{Replicating Cat States}

We would like to generate a superposition of coherent states (cat states
\cite{NN06}-\cite{O07a})
\[
\Psi = \sum_k \gamma_k \vert \beta_k \rangle ,
\]
with $\sum_{k,l} \gamma_k^\ast \gamma_l e^{\langle \beta_k , \beta_l \rangle }=1$ for
normalization. It is easy to see that pairs of coherent states are compatible in the sense of the superposition principle.

We know that the modulator (with fixed structure $L_M = \lambda (t) a , H_M = \omega (t) a^\ast a$) prepared in coherent state
$\vert \alpha_k \rangle$ will replicate the open system with coherent state $\vert \beta_k \rangle$ for the

\begin{eqnarray}
 \beta_k \left( t\right) = \xi \left( t\right) \alpha_k .
\label{eq:coherent_state_condition1}
\end{eqnarray}

The principle of superposition therefore implies that the initial state $\sum_k \gamma_k \vert \alpha_k \rangle$ for the modulator
will then replicate the cat state $\Psi = \sum_k \gamma_k \vert \beta_k \rangle$. Note that the $\beta_k$ that may be generated this way
must take the form (\ref{eq:coherent_state_condition1}). This is somewhat restrictive since we cannot obtain independent pulse, only pulse which differ by
the scale factors $\alpha_k$. However this already gives a wide class of cat states for practical purposes.

\begin{figure}[h]
 \centering
\includegraphics[width=0.40\textwidth]{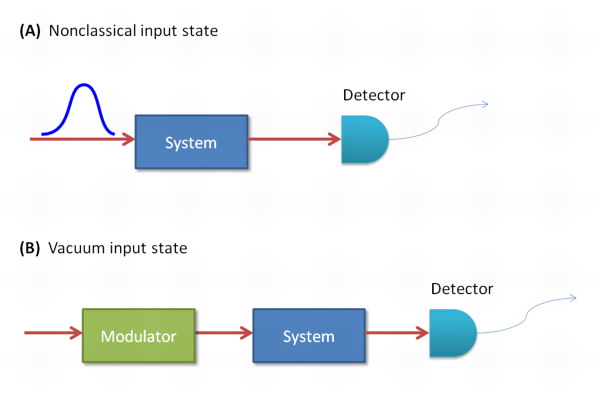}
  \caption{(color online) Continuous measurement of (A) the output of a system driven by non-classical noise, and (B) and equivalent
  model from a modulated vacuum noise.}
  \label{Fig:measure}
\end{figure}

\section{Conclusion}
We extended the classical concept of a modulating filter, which colors white noise, to the quantum domain. The usefulness of this concept is that systems driven by nonclassical states (which may be difficult to analyze directly using quantum stochastic techniques) may be described by an equivalent model with vacuum white noise input. Here one replaces the problem with a cascaded model of modulator and system, with the modulator processing vacuum noise and the colored noise is then fed into the system. We have shown that certain modulator models replicate the original system and noise model in a strong sense: that is we show that general quantum stochastic integral processes on the system plus noise space have identical averages in the original model with non-classical input and in the modulated model with vacuum input. This in particular establishes equivalence of the system dynamics (effectively the same Ehrenfest equations for system observables) as well as equivalence of the outputs. The latter point is of major importance with regards to quantum trajectories (quantum filtering theory) since whenever we perform continuous measurements (e.g., quadrature homodyne, or photon counting) on the output we have that the measurement processes of the original model and the modulate model are statistically identical. See Figure \ref{Fig:measure}.

As a conceptual tool, this opens up the prospect of extending known results on quantum trajectories for vacuum inputs to models with nonclassical inputs, see for 
\cite{GJNC12,SZX}.
One such feature which we will address in a future publication is the issue of filter convergence, that is, when does the estimated conditional density operator
converge to the true conditional density operator when one starts with the wrong initial state $\psi_0$ for the system - this has been treated for vacuum inputs \cite{RvH_IDAQP,Rouchon}, but is largely unknown in the case of non-classical inputs.

\textbf{Acknowledgement:} The authors acknowledge support through the Royal
Academy of Engineering UK and China scheme and EPSRC project EP/L006111/1.

\thinspace


\begin{thebibliography}{99}
\bibitem{L01}  A.I. Lvovsky, \emph{et al.,}
% H. Hansen, T. Aichele, O. Benson, J. Mlynek, and S. Schiller,
Phys. Rev. Lett. \textbf{87}, 050402 (2001)

\bibitem{K}  A. Kuhn, M. Hennrich, and G. Rempe, Phys. Rev. Lett. \textbf{89}, 067901 (2002)


\bibitem{Y02}  Z. Yuan, B.E. Kardynal, \emph{et al.,}
%R. M. Stevenson, A. J. Shields, C. J. Lobo, K. Cooper, N. S. Beattie, D. A. Ritchie, and M. Pepper,
Science \textbf{295}, 102 (2002)

\bibitem{K04}  J. McKeever, A. Boca, \emph{et al.,}
% A. D. Boozer, R. Miller, J. R. Buck, A. Kuzmich, and H. J. Kimble,
Science \textbf{303}, 1992 (2004)

\bibitem{E11}  C. Eichler, D. Bozyigit, C. Lang, L. Steffen, J. Fink, and A.
Wallraff, Phys. Rev. Lett. \textbf{106}, 220503 (2011)

\bibitem{NN06}  J. S. Neergaard-Nielsen, %\emph{et al.}
B.M. Nielsen, C. Hettich, K. M\o lmer, and E. S. Polzik,
Phys. Rev. Lett. \textbf{97}, 083604 (2006).

\bibitem{O07}  A. Ourjoumtsev, R. Tualle-Brouri, J. Laurat, and P. Grangier,
Science 312, 83 (2006).

\bibitem{O07a}  A. Ourjoumtsev, H. Jeong, R. Tualle-Brouri, and P. Grangier,
Nature 448, 784 (2007).

\bibitem{H02}  G.S. Vasilev, D. Ljunggren, A. Kuhn, New Journal of Physics, \textbf{12}, 063024 (2010)
(2001)

\bibitem{Kun} P.B.R. Nisbet-Jones, J. Dilley, D. Ljunggren and A. Kuhn, New Journal of Physics,  \textbf{13}, 103036 (2011)

\bibitem{R03}  T.C. Ralph, A. Gilchrist, and G.J. Milburn, W.J. Munro, S. Glancy, Phys. Rev. A
\textbf{68}, 042319 (2003)

\bibitem{G02}  N. Gisin, G. Ribordy, W. Tittel, and H. Zbinden, Reviews of
Modern Physics, \textbf{74}, 145 (2002)

\bibitem{GJNC12}  J.E. Gough, M.R. James, H.I. Nurdin, J. Combes, Phys. Rev.
A \textbf{86}, 043819 (2012)

\bibitem{SZX} H. Song, G. Zhang, Z. Xi,  arXiv:1307.7367

\bibitem{C97}  J.I. Cirac, P. Zoller, H.J. Kimble, and H. Mabuchi, Phys.
Rev. Lett. \textbf{78}, 3221 (1997).

\bibitem{HP}  R.L. Hudson and K.R. Parthasarathy, Commun. Math. Phys.
\textbf{93}, 301-323 (1984)

\bibitem{EH88}  M.P. Evans, R.L. Hudson, Springer LNM 1303, 69-88 (1988)

\bibitem{QFN1}  J.E. Gough, M.R. James, Commun. Math. Phys. \textbf{287},
1109-1132 (2009)

\bibitem{GJSeries}  J.E. Gough, M.R. James, IEEE Trans. Autom. Control
\textbf{54}, 2530 (2009)

%\bibitem{Partha}  K.R. Parthasarathy, An Introduction to Quantum Stochastic Calculus, Birkhauser, Berlin, (1992)

\bibitem{GK10}  J.E. Gough, C. Koestler, Commun. Stoch. Analysis, \textbf{4}, no. 4, 505-521 (2010)

\bibitem{RvH_IDAQP} R. van Handel, Infin. Dimens. Anal. Quantum Probab. Relat. Top. \textbf{12}, 153-172 (2009)

\bibitem{Rouchon} P. Rouchon, IEEE Trans. Automat. Contr., \textbf{56},  Issue 11, 2743 - 2747 (2011)

\end{thebibliography}
\end{document}